\documentclass[conference]{IEEEtran}

\usepackage{times}
\usepackage{helvet}
\usepackage{courier}
\usepackage{graphicx}
\usepackage{subcaption}
\usepackage{url}
\usepackage[utf8]{inputenc}
\usepackage[margin=0.7in]{geometry}

\newcommand{\etal}{\textit{et al.}}
\renewcommand{\footnotesize}{\fontsize{8pt}{10pt}\selectfont}
\title{Towards a Resilient Machine Learning Classifier\\- a Case Study of Ransomware Detection
}

\author{Chih-Yuan Yang and Ravi Sahita\\
Security and Privacy Research, Intel Labs, Hillsboro, Oregon, USA }
\date{October 2019}

\begin{document}
\maketitle

\begin{abstract}
The damage caused by crypto-ransomware, due to encryption, is difficult to revert and cause data losses. In this paper, a machine learning (ML) classifier was built to early detect ransomware (called crypto-ransomware) that uses cryptography by program behavior. If a signature-based detection was missed, a behavior-based detector can be the last line of defense to detect and contain the damages. We find that input/output activities of ransomware and the file-content entropy are unique traits to detect crypto-ransomware. A deep-learning (DL) classifier can detect ransomware with a high accuracy and a low false positive rate. We conduct an adversarial research against the models generated. We use simulated ransomware programs to launch a gray-box analysis to probe the weakness of ML classifiers to improve model robustness. In addition to accuracy and resiliency, trustworthiness is the other key criteria for a quality detector. Making sure that the correct information was used for inference is important for a security application. The Integrated Gradient method was used to explain the deep learning model and also to reveal why false negatives evade the detection. The approaches to build and to evaluate a real-world detector were demonstrated and discussed.
\footnote{This paper was presented at the Conference on Applied Machine Learning for Information Security 2019, Washington DC}

\end{abstract}

\providecommand{\keywords}[1]
{
  \small	
  \textbf{\textit{Keywords---}} #1
}


\section{Introduction}

Ransomware is a type of malware which hijacks user’s resource or machine and demands for a ransom. It was estimated to cost business more than \$75 billion in 2019 and continues to be a problem for enterprises \cite{ransomfacts}. Ransomware can be divided into two main categories, the locker- and the crypto- ransomware \cite{Al-rimy2018}. The locker-ransomware hijacks resource without using the encryption, but crypto-ransomware does. Due to the encryption, the file encrypted by the crypto-ransomware in most cases is difficult to revert or decrypt. Even with a proper backup, there is still a chance to miss partial data between ransomware strike and the last backup. An endpoint protection software based on binary signature may not be able to block an unseen ransomware. The behavior-based detection \cite{dynaAnalysis, unveil} combined with a proper backup mechanism was proposed to be one of mitigation solutions. 

\begin{figure}[h]
\centering
\includegraphics[width=0.4\textwidth]{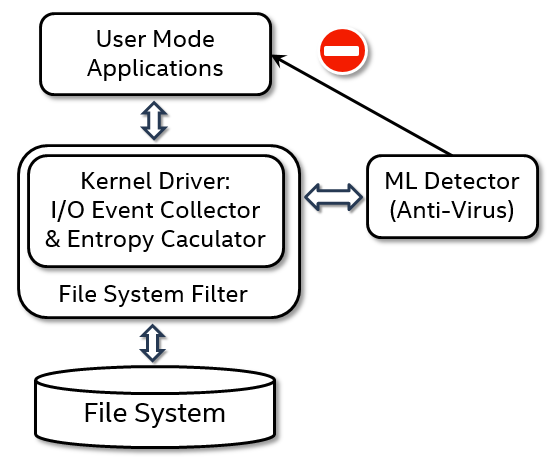}
\caption{\label{fig:design} The purpose of detector is to detect crypto-ransomware early and contain the damage. The I/O event collector could be a kernel driver, a file system filter, or an application. The ML detector can be part of Anti-virus program which can stop or kill positive processes.} 
\end{figure}

In this paper, machine learning (ML) and deep learning (DL) classifiers were proposed to early detect the crypto-ransomware based on its behaviors. These classifiers can monitor the pattern of input/output (I/O) activities and can minimize the damages by an early detection. The detector could be a part of an endpoint protection application and help to find a new ransomware if static-based detection can't catch it (Figure \ref{fig:design}). Although few files may get encrypted before the detection, the dynamic-based classifier would be still valuable, if most of the data can be saved for an enterprise user with lots of data in shared drives. 

To collect the behavior data, the ransomwares was executed in a Windows sandbox system and their file I/O activities were logged \cite{gan}. The time-series data was analyzed by the DL algorithm, long short term memory (LSTM), and ML algorithm, N-gram featured linear support vector machine (SVM). We found that a naive trained classifier even with good accuracy (\textasciitilde98\%) and low false positive rate (\textasciitilde1-3\%)) didn't perform well at real-world deployment. Issues include: 1. Ransomware can't be detected early; 2. The accuracy is sensitive to size of sliding window and 3. False alarms from some applications etc. 

The bootstrapping method by Bradley Efron \cite{bootstrapping} is a well-known technique used to estimate the properties of a statistic. The subclass of bootstrap methods, the block bootstrap, were found effective to improve the performance of classification in our experiment. This over-sampling mechanism can generate data similar to the samples used for the online detector. The three issues mention above can be mitigated after model retraining by the augmented dataset and by adding a new dimension to the featuring method.

For security application, it is important to ensure that the ML classifiers making decisions base on meaningful features. Interpreting machine learning model become a needed process to elucidate important aspects of learned models and to ensure the reliability. Several saliency methods for the image model explanation were verified by Adebayo, J \etal \cite{smap}. A good saliency method should satisfy two fundamental axioms: sensitivity and implementation invariance. The Integrated Gradient (IG) \cite{ig_aa} method was selected for our case, because it fulfills the axioms and is easy to implement without the needs to retrain or to instrument ML models. IG addresses the "gradient saturation" by summing over scaled versions of inputs. The integrated gradient along the \(i^{th}\) dimension for an input \(x\) and baseline \(\bar x\) is defined as  
\begin{equation}
    IG_{i}(x)=(x_{i}-\bar x_{i})\times \int_{\alpha=0}^1\frac{\partial F(\bar x + \alpha(x-\bar x))}{\partial x_{i}}d\alpha 
\end{equation}
where \(\frac{\partial F(x)}{\partial x_{i}}\) is the gradient of \(F(x)\) along the \(i^{th}\) dimension. The baseline input \(\bar x\) that represents the absence of a feature in the original input \(x\).

In our experiment, the attribution of each time step in LSTM model can be generated. We observed that the attribution pattern did match the known malicious I/O activities. Then the fidelity of classifier can be verified and confirmed. We also apply the IG method to explain how an adversarial sample bypasses a detection. 

A ML/DL model without adversarial mitigation may be vulnerable to adversarial attacks \cite{goodfellow}. A simulated ransomware, the Red team, was developed to probe the blind spots of our classifiers. This simulated program can perform the core ransomware behaviors, e.g. the malicious encryption, and also  configurable benign I/O activities, e.g. file creation etc. With minor change to the I/O behavior of encryption, the Red team found no difficulty to bypass the ML detection. We conclude that an adversarial mitigation is necessary procedure to fortify the ML/DL classifier especially when dataset size is limited or the featuring is simple. 

The simulated adversarial program was found very helpful to fortify the model. It not only discloses the weakness of the model, but also serves as an adversarial sample generator. In addition to the regular ML/DL training-testing iteration for model optimization, we emphasize the adversarial training iteration. The real adversarial samples by the polymorphic Red team were collected to augment the dataset. Combining with data bootstrapping and model explanation techniques, the resiliency and fidelity of the model can be enhanced and ensured. The tips and lessons learned for each steps of two-iteration pipeline will be discussed in the result session. We believe this in-depth analysis can be a general recommendation for ML/DL application in cybersecurity field. 

Data Augmentation is a popular solution to enhance the size and quality of training dataset for deep learning models \cite{data_aug_srv}. For time series data, dataset augmentation can be done by data warping, synthetic minority over-sampling technique (SMOTE) \cite{smote}, or generative neural networks. The “data warping” generates new samples by transformation on raw data (such as image pixel) directly. It includes the affine transformation (translation, shearing, rotation, flip) and elastic distortions (scaled normalized displacement in image space) etc. \cite{aug_warp} The “SMOTE” works on the feature space. New samples were generated by randomly selection of real features from the minority dataset. The generative neural network is a powerful method to learn the internal data representation and then to generate new data samples with some variations. The Variational Autoencoder (VAE) \cite{vae} and Generative Adversarial Networks (GAN) \cite{gan} are two popular ones.

The samples generated by SMOTE were found to be very similar to existing samples. It may be easy to cause overfitting, if too many “similar” synthesized samples in the dataset. Depending on total number of time step and feature count, the GAN and VAE may need a bulky network architecture and may be difficult to train. In this paper, a simple and fast data augmentation, the "keyed" method, was proposed to synthesize time series samples and then to prob and to help to understand the DL models. \\

In summary, this paper makes the following contributions:
\begin{itemize}
  \item A general sandbox system was built for evasive malware analysis. This system include two parts: a bare-metal machine and a user activity simulator. The fundamental solution to catch anti-VM or anti-sandbox malware is using a bare-metal system. By utilizing the dual boot support in Linux, a bare-metal sandbox can be refreshed quickly by copying a disk image. The user activity simulator is important to trigger malware which spying mouse movement or keyboard strokes etc. before starting its malicious intents. The activity simulator can make a sandbox like an active regular user machine.
  \item A full stack of ML/DL development process was demonstrated. The real-world issues of ML detector were discovered by an online detector. The processes from performance improvement to adversarial analysis  were revealed. Also the "keyed" data augmentation method was developed to prob the deep learning model. And the model was explained by the Integrated Gradient method.
  \item The Red team, a simulated ransomware program, was proofed to be effective for improving the resiliency of ML detector. It can help to find the weakness of ML classifier and also synthesize false-negative samples to augment the dataset. The model robustness can be enhanced after few iterations of adversarial re-training.  
\end{itemize}

\section{DataSet}
\subsection{Crypto-ransomware}
\begin{figure}[h]
\centering
\includegraphics[width=0.45\textwidth]{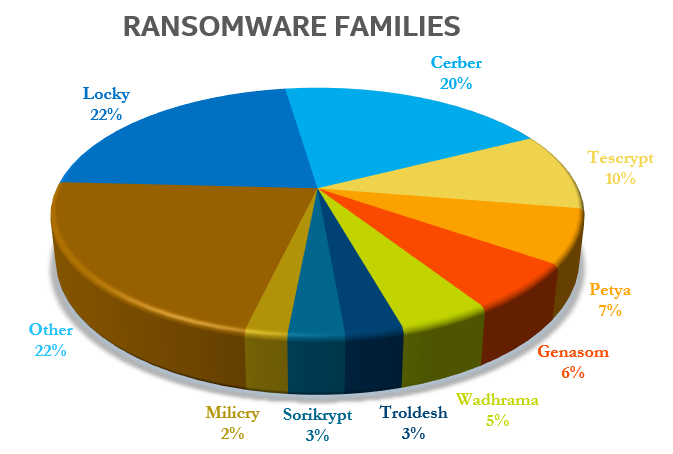}
\caption{\label{fig:families} The ransomware family distribution in our dataset. The distribution is based on the Microsoft\textsuperscript{\textregistered}'s label provided by VirusTotal\textsuperscript{\textregistered}.} 
\end{figure}

\textasciitilde22k of Ransomware binaries, Windows executable, were downloaded from VirusTotal\textsuperscript{\textregistered} based on Microsoft\textsuperscript{\textregistered} and Kaspersky\textsuperscript{\textregistered}'s label, "ransom". \footnote{Intel and the Intel logo are trademarks of Intel Corporation in the U.S. and/or other countries. Other names and brands may be claimed as the property of others.} Each binary was executed for 5 minutes in a bare-metal sandbox system. The sandbox is a regular PC machine running Windows 8.1 without any virtualization. To activate more ransomware, an automated user activities simulator by AutoIt \cite{autoit} was running during the ransomware execution. The AutoIt is a scripting language designed for automating the Windows GUI activities. It can simulate keystrokes, mouse movements and window/control manipulations. Ransomware will run as an administrator with full access to the sandbox except the folder for log collection. Also each sandbox was loaded with decoy or canary files in C: drive, Documents, Downloads, Music, Pictures and Videos folders etc. These files were designed to be the targets of crypto-ransomware. At the end, we identify around 4.4k of active ransomware by checking the existence of the events from decoy files or folders. The distribution of active ransomware family is shown in Figure \ref{fig:families}. The dataset has various of ransomware families.

\subsection{Bare-Metal System and Sandbox Farm}

\begin{figure}[h]
\centering
\includegraphics[width=0.5\textwidth]{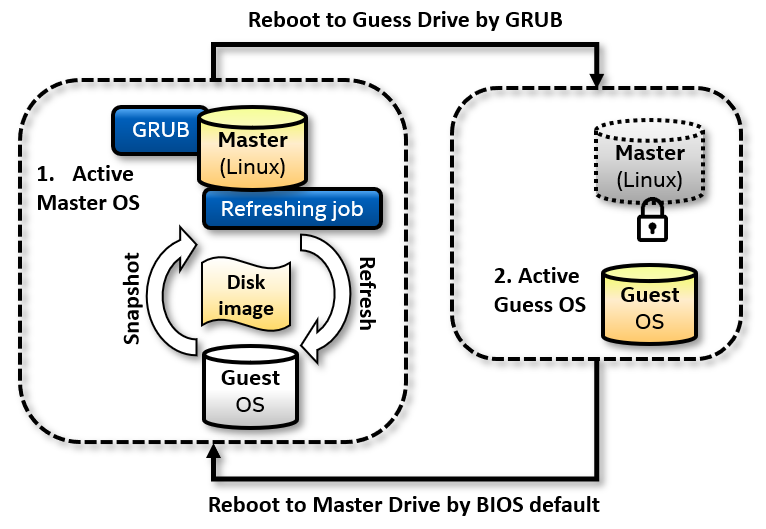}
\caption{\label{fig:sandbox} The life cycle of bare-metal machine by alternately boot between two physical drives. The dash-line boxes represent the states of bare-metal machine: Active “Master OS” and Active “Guest OS”. The Master OS utilizes the boot loader supporting multiple OS to boot to the Guest OS. When the Guest OS reboot, it will boot to the default drive defined in BIOS. The Master drive was protected when booting to Guest OS. The Guest drive could be refreshed or saved/snapshot-ed in active Master OS.} 
\end{figure}

\begin{figure}[h]
\centering
\includegraphics[width=0.5\textwidth]{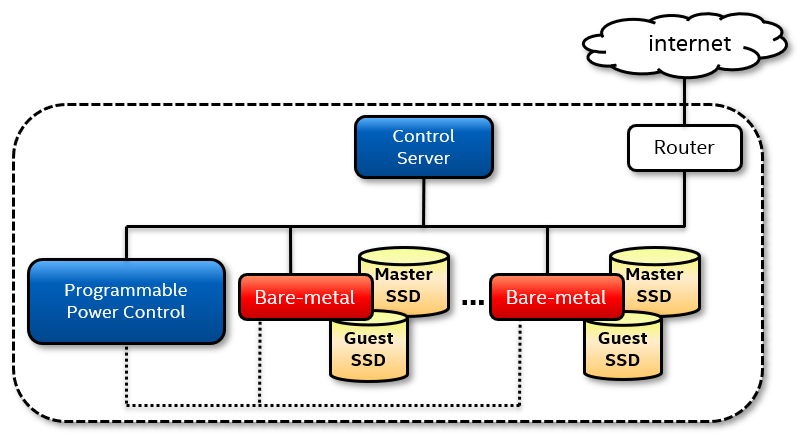}
\caption{\label{fig:farm} The farm of bare-metal sandbox. The Control server can dispatch tasks to bare-metal machines, collect the results and monitor the status. If machine hung, the server can trigger the programmable power control to re-start the machine.} 
\end{figure}

A fast bare-metal sandbox system is built to avoid using virtual machine (VM). It can be easily implemented without error-prone modifications on operating system or storage device firmware. It supports any guest operating system (OS) on any machine with fast storage devices, e.g. M.2 SSD. The guest OS can quickly be refreshed and be automated for a continuously sandboxing process. The details of sandbox life cycle can be found in Figure \ref{fig:sandbox}. The Master drive can be hided and protected by three-fold mechanisms: 1. A trigger listens to the “Plug and Play” (PnP) event to remove the Master drive devise (if Guest OS supports the trigger). 2. A custom service/cron job polling the existence of Master device for removal if found. 3. A simple kernel driver in Guest OS to remove the Master drive device if found. The Figure \ref{fig:farm} shows the architecture of a sandbox farm. The malicious binary execution and data collection was fully automated and is good for a large-scale malware analysis. The Control server and the programmable power control are two key components to orchestra the sandbox farm.

\subsection{Behavior Data Collection}

The behavior data was collected by a proof-of-concept (POC) application which implemented the C\# .Net API FileSystemWatcher (FSW). Whenever there is a file input/output (I/O) activities, the callback function will be invoked and the I/O activities can be logged. The execution log is in a CSV format and contains timestamp, event type and the entropy of target files. The entropy of target file was calculated by Shannon entropy (\(H\)) method \cite{entropy}.

\begin{equation}
    H = - \sum_{i=1}^{n} p_i log_2 (p_i)
\end{equation}
where \(p_i\) is the probability of occurrence of the \(i^{th}\) possible value of the source data.

\begin{figure}[h]
\centering
\includegraphics[width=.5\textwidth]{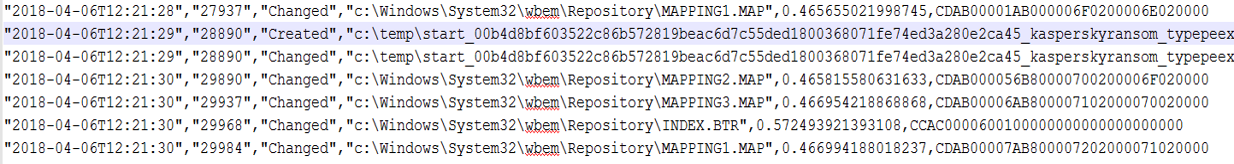} 
\caption{\label{fig:sample_csv} A sample execution log collected by behavior data collector. It has fields: time stamp, event name, targeted filename, entropy etc. in CSV format} 
\end{figure}

When a file was encrypted, the normalized Shannon entropy will be high and close to value 1.0. A regular file may have an entropy around 0.5. To avoid overload the CPU, the entropy was calculated only from the first 1 Mb of the file. The sample I/O event log can be found at Figure \ref{fig:sample_csv}.

\subsection{Data Block Bootstrapping and Augmentation}

\begin{figure}[h]
\centering
\includegraphics[width=.5\textwidth]{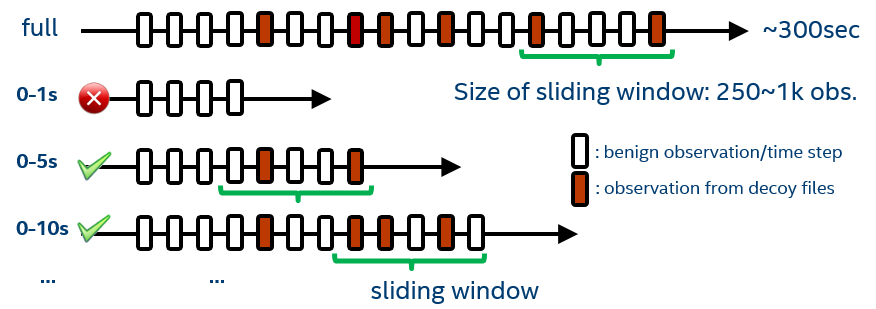}
\caption{\label{fig:data_aug_exp} The data bootstrapping and augmentation. The first row is simplified diagram for the full execution log. Each box is one time step or an I/O event. Early-stage samples were sliced from each sample by different time period and then filtered by decoy events. The sliding-window samples were further sliced from the end of early-stage samples by different window sizes.} 
\end{figure}

The data bootstrapping was done by two methods: The early part of execution log was sliced by different time periods. The length of time period are 0\textasciitilde1 sec; 0\textasciitilde5sec; 0\textasciitilde10sec and up to 0\textasciitilde160sec. The Figure \ref{fig:data_aug_exp} illustrates the slicing process in details. The intuition to slice early part of execution log is to have a dataset focus on early stage of ransomware behavior. 

The second set of sample was created by further slicing the early-stage samples with different sizes of sliding windows. Sliding-window samples have either 250, 500 or 1,000 time steps. The design of sliding-window sample is based on the usage of online detector. The detector inferences a sample extracted from I/O event stream by a sliding window method. 

The benign dataset was collected from \textasciitilde100 benign-ware running in the same sandbox or from regular Windows machines in a normal office.

\section{Methods}
\subsection{Model Architecture}
Around 3.7k of ransomware execution log was allocated with a similar amount of benign sample for a supervised ML training. The I/O event and file entropy were used for building the features for ML training. Entropy was bucketed by an arbitrary selected range from 0(-), 0.2+, 0.4+, 0.6+, 0.8+, 0.9+. The raw data from execution log then was categorized into few distinct numbers (features) and then further encoding by ML algorithms.

The deep learning algorithms, long-short term memory (LSTM) and the regularized linear support vector machine (SVM) with bag of N-gram, N=1 and 2, were used to train models\cite{mlaas}. The LSTM is a popular recurrent neural network (RNN) which can catch long term dependence of time data. The architecture of LSTM contains a one-hot encoding layer, a LSTM layer with 16 nodes, a 50\% dropout and a softmax layer. The linear SVM with N-gram algorithm is a powerful but light-weight method for analyzing sequence data.  ML libraries used include Keras\cite{keras} with Tensorflow backend\cite{tensorflow}, Scikit-Learn\cite{sk} or Spark MLlib\cite{spark}. An open source machine learning as-a service \cite{mlaas}, a big data based platform, makes the ML analysis very handy.

\subsection{Online Detector}
The POC detector application was developed by Python which utilized the ML model to detect ransomware pattern from I/O event stream. It can continuously extract a sequence of I/O events by various size of sliding window and make a real time prediction. It depends on the Python libraries, e.g. Keras\cite{keras}/Tensorflow\cite{tensorflow} and Scikit Learn\cite{sk}.

\subsection{Integrated Gradients}
The implementation of Integrated Gradients can be downloaded from GitHub site \cite{ig}. The calculation for attribution is straightforward and quick by Python. The only criteria is that each layer in ML model needs to have a gradient. We use zero for the baseline inputs with 50 as the number of steps.

\subsection{Adversarial Studies by the Red Team and the "Keyed" Augmentation}
The simulated ransomware is a C\# application which can do the core business of ransomware, e.g. rename, delete or encrypt files etc. The grey box attack was implemented. It attacks the ML model by assuming features based on I/O events and file entropy without knowing the ML model architectures, algorithms or weights. The evasive tricks have two categories: changing the tempo of I/O activity or the entropy of encrypted file. The tempo variants can be done by actively inserting benign I/O events between malicious I/O events or slowing down the malicious activities by a sleep. The benign events will be inserted from other process during the sleep. Changing the file entropy can be done by inserting dummy data into output files or encrypt selected regions of file. The high entropy can be lowered by dummy low-entropy data or partial encryption.

\begin{figure}[h]
\centering
\includegraphics[width=.5\textwidth]{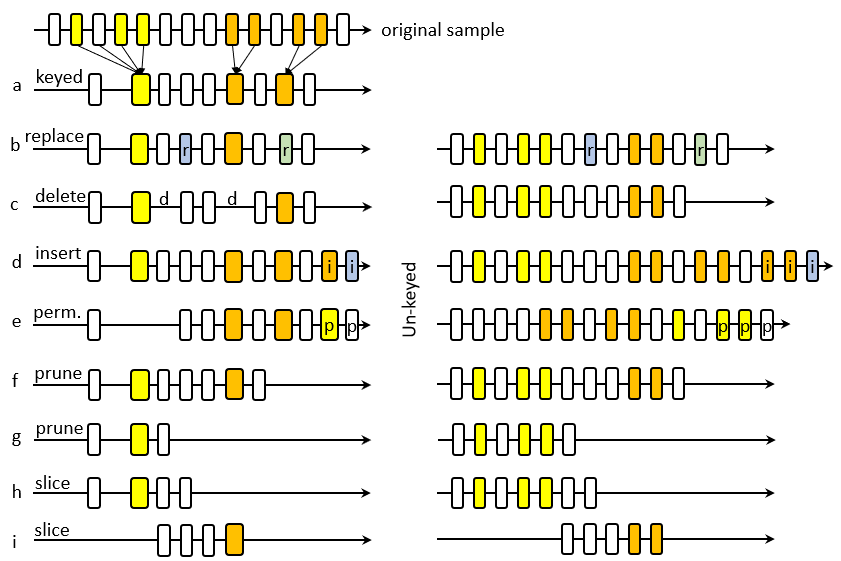}
\caption{\label{fig:data_aug} "Keyed" Data Augmentation Methods. Each box represent an observation/time step. Left side of diagram represent the “keyed” sample. Right side is “Un-keyed” samples. New samples (b-i) were generated based on the keyed sample (a).  The “key” and “key group” observations (orange and yellow boxes) were identified by human. (b) Some observations were replaced. (c) Some observations were deleted. (d) New observations were inserted. (e) Observations were permuted.  (f \& g) Few observations was removed or pruned. (h \& i) Few observations was extracted (sliding window). } 
\end{figure}

The "keyed" data augmentation method for adversarial study was developed (Figure \ref{fig:data_aug}). With the help of domain experts, a set of “key observations” from real samples were identified based on the golden rules or well-known heuristics. In our case, the key observation are events from decoy files or folders. The purpose of “key observations” is to find out the important time steps which can represent the “key” traits of real samples. If the order of multiple key observations is critical, these observations should be groups into a “key group” which can be treated as a single building block of synthesis process in Figure \ref{fig:data_aug}(a). The “key group” can preserve the ordering and completeness of related observations. Each synthesized sample should have at least one “key” or “key group” observation in it. Samples generated are based on the “keyed” real samples. The “key” or “key groups” were operated as a single unit in synthesis process (Figure \ref{fig:data_aug}a). The observation or key group can be seen as a “block” which can be either replaced, removed, inserted or permuted, see Figure \ref{fig:data_aug}(b-e). A series of blocks can be sliced or pruned, Figure \ref{fig:data_aug}(f-i).

\section{Results}
\subsection{Initial Model And Issues}
\begin{figure}[h]
\centering
\includegraphics[width=.5\textwidth]{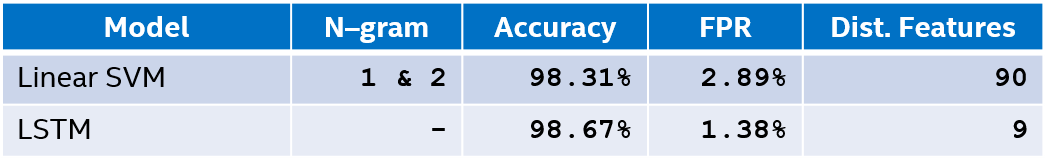}
\caption{\label{fig:result_orig} The result of naive trained ML model. LSTM has 9 distinct features: Create, Rename, Delete and six Change events. The six Change events were from 6 buckets of content entropy. Linear SVM with N-gram has 90 features: 9 from N=1 and 9x9 from N=2} 
\end{figure}

The classifier was trained by a regular ML pipeline on a big-data platform \cite{mlaas}. The initial result of both ML models have similar accuracy, around 98\%, and false positive rate (FPR), around 2\%.  from Figure \ref{fig:result_orig}. 

Then the detector application by this ML classifier was tested in a live settings. Even the classifier has a good accuracy during the training, three majors issues were found. 1. False alarms were triggered from some applications, 2. small size of sliding window missed some ransomware and  3. Ransomware can't be detected early. The naive trained classifier didn't work well in the real world condition.

\subsection{Early Detection and Size of Sliding Window Issues}

\begin{figure}[h]
\centering

\begin{subfigure}[b]{0.5\textwidth}
   \includegraphics[width=1\linewidth]{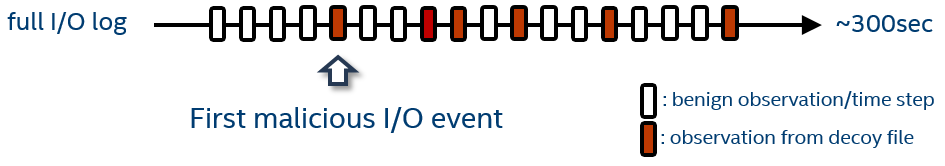}
   \caption{}
   \label{fig:Ng1} 
\end{subfigure}

\begin{subfigure}[b]{0.5\textwidth}
   \includegraphics[width=1\linewidth]{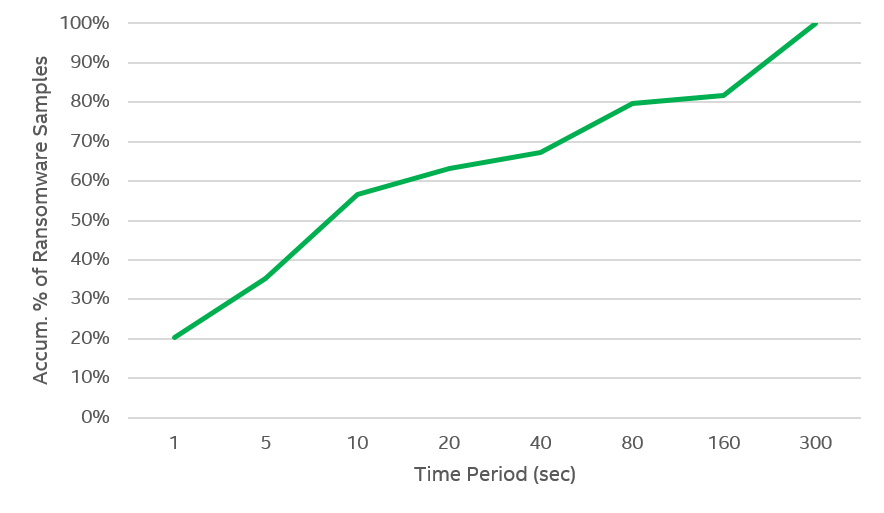}
   \caption{}
   \label{fig:Ng2}
\end{subfigure}

\caption{\label{fig:starting_time} (a) The rough starting time can be found by the first event from decoy files. (b) The accumulated percentage of ransomware with decoy events vs. different time period of execution } 
\end{figure}

To fix these three issues, we first studied the starting time of ransomware. Based on the time region of decoy events, Figure \ref{fig:starting_time}(a), the approximate starting time region of malicious activity can be identified. The starting time for the dataset were found to have a distribution along the span of execution time \ref{fig:starting_time}(b) . Around 20\% of ransomware shows malicious events in first second of execution. However, around 20\% of them didn't show any malicious activities until very late of execution (\textgreater160sec). This distribution of starting time will make the dataset preparation for early detection a bit tricky.

\begin{figure*}[ht]
\centering
\begin{subfigure}[b]{0.8\linewidth}
   \includegraphics[width=1\linewidth]{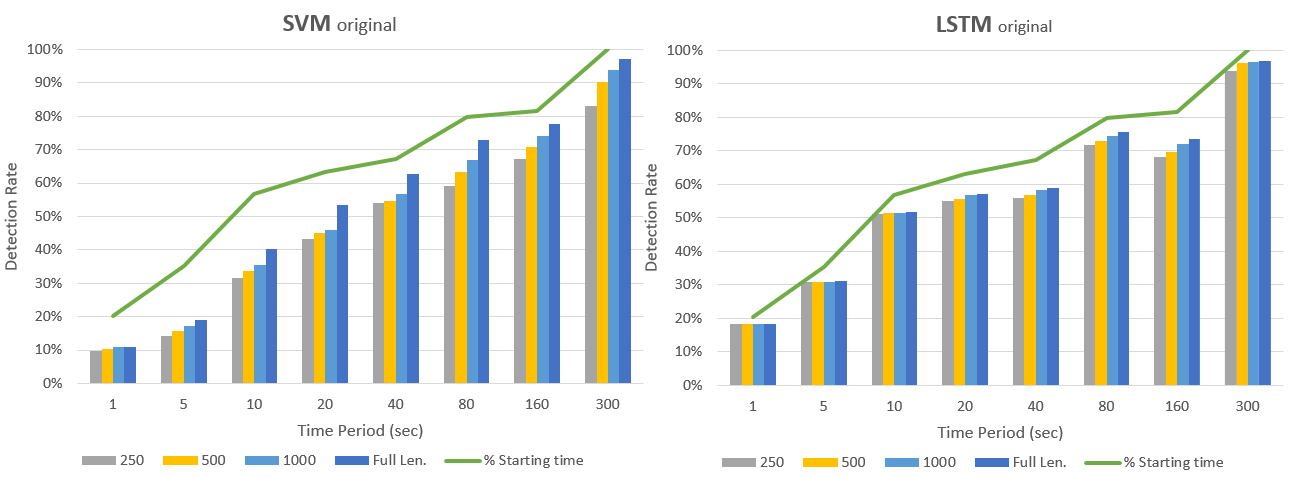}
   \caption{Results of the original ML model}
   \label{fig:ew1} 
\end{subfigure}
\begin{subfigure}[b]{0.8\linewidth}
   \includegraphics[width=1\linewidth]{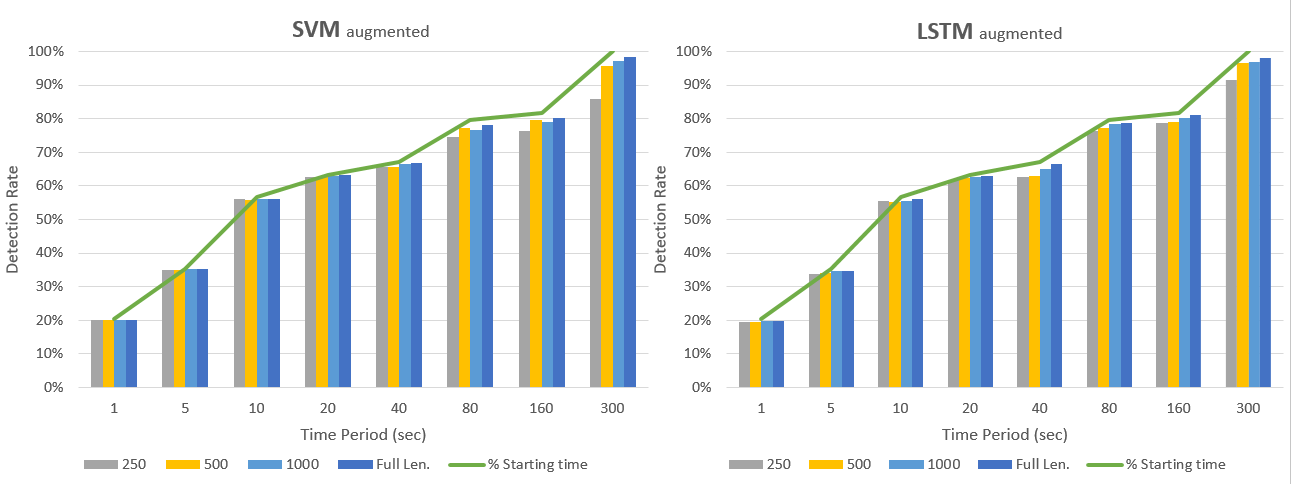}
   \caption{Results of the ML model trained by augmented dataset}
   \label{fig:ew2}
\end{subfigure}
\caption{\label{fig:early_wind_test} The detection rates of early-stage and sliding-window samples are in colored bars. The green line is the percentage of ransomware in each execution time period. It is the same result of Figure \ref{fig:starting_time}(b). When a ML model has a good performance, the top of bar should be close to the green line and the detection rate for different size of sliding window should be similar. } 
\end{figure*}

To measure the performance gap, early-stage samples and sliding-windows samples were synthesized from \textasciitilde\textit{700 out-of-sample} ransomware log. These two set of samples were predicted by the ML classifier. The performance results can be found at Figure \ref{fig:early_wind_test}(a). 

The detection results for original SVM and LSTM models can be found in the Figure \ref{fig:early_wind_test}(a). The gap between green line and color bars means some ransomware samples were missed by the model. The early-stage samples for SVM model, e.g. X-axis time period = 1, 5, 10 sec etc., have most prominent gaps. Also when sliding window is getting smaller, the detection rate is getting worse. We suspected the size effect of sliding window is related to the available amount of data. And the cause of the early-stage issue could be that the classifier didn't focus on the early stage region for decision making. Comparing two models, the LSTM model has a smaller gap and is less sensitive to the sizes of sliding window.

To build a model focus on a intended region of execution log, the dataset was augmented by the bootstrapping. Early-stage and sliding-window samples were generated from each execution log in the original training dataset. Generated samples without any decoy events were be excluded to make sure its maliciousness, Figure \ref{fig:data_aug_exp}. The sample count of dataset was increased to around 17.2k. After a re-training, the result of re-test is in Figure  \ref{fig:early_wind_test}(b). Both the early-stage and sliding-window issues were fixed or minimized for both LSTM and SVM models. 

\subsection{False Alarm Issue}
\begin{figure}[h]
\centering
\includegraphics[width=.5\textwidth]{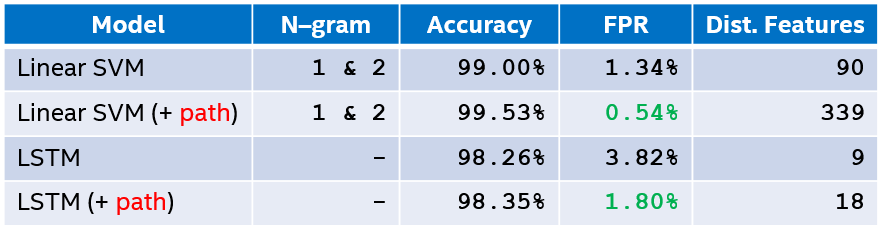}
\caption{\label{fig:result_pathflag} The result of path-flagged ML model. The green numbers in FPR column do have a lower number comparing to the model without the path flags } 
\end{figure}

By inspecting few false positive samples, we found some benign-wares show behaviors similar to ransomware e.g. delete or rename many files or modify files with high entropy. To increase the resolution of classifier, we added a new dimension, the "system path" flag, to the feature. The system paths were white listed, e.g. c:\textbackslash{}Windows, c:\textbackslash{}ProgramData, c:\textbackslash{}Program Files, c:\textbackslash{}Progra\textasciitilde{}, c:\textbackslash{}AppData etc. Most of the benign-ware works on system folders, but ransomware likely works on non-system folders. The output of retraining model is at Figure \ref{fig:result_pathflag}. The FPR is lower for path-flagged model. Also around 22k out-of-sample clean execution logs were tested by this path-flagged model. The FPR was lower from 0.15\% to 0\% for SVM and 0.09\% to 0.04\% for LSTM model. We believe the path flag does help to minimize the false alarm. 

\subsection{Model Fidelity}
\begin{figure}[h]
\begin{subfigure}[b]{1\linewidth}
   \includegraphics[width=1\linewidth]{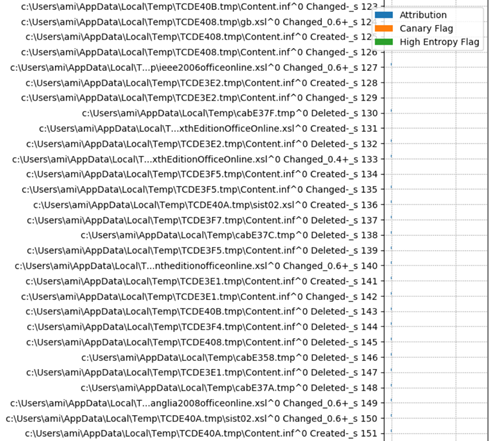}
   \caption{The attribution of \textbf{clean region} of ransomware execution log. All of the time steps in this plot have negative or zero attribution. \\
   }
   \label{fig:ig1} 
\end{subfigure}

\begin{subfigure}[b]{1\linewidth}
   \includegraphics[width=1\linewidth]{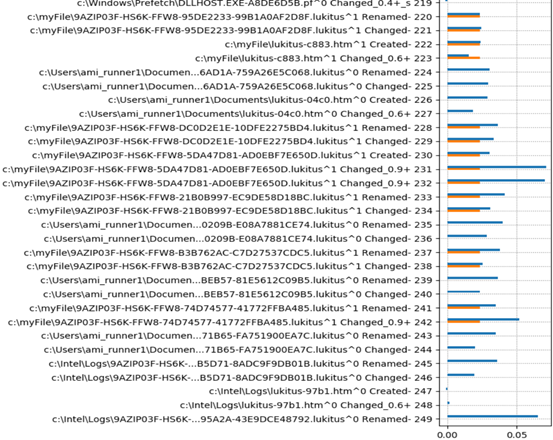}
   \caption{The attribution of \textbf{malicious region} of ransomware execution log. The blue bar point to right has positive contribution to positive prediction. The orange bar is a label for the events from decoy files (found by the keyword "myFile"). The pattern for blue and orange bars are matched well.
   }
   \label{fig:ig2}
\end{subfigure}

\caption{\label{fig:ig} The attribution results of Integrated Gradients for a ransomware execution log. The left part of the diagram is the list of raw features (filename, event type, entropy \& sequence number). The right part is the attribution amount from Integrated Gradients method (blue bar) and the flags for decoy/canary files (orange bar) } 
\end{figure}

The model fidelity for LSTM was verified by Integrated Gradients. The attribution of ransomware sample can be found in Figure \ref{fig:ig}. The blue bars are the attribution amount calculated by Integrated Gradients. The orange bars are flags to label events from decoy or canary files. They can reveal the malicious region of execution log. In the clean region, Figure \ref{fig:ig}(a), there is no decoy events and attributions are all closed to zero for all time steps. However, in the region with clustered orange bars, the Figure  \ref{fig:ig}(b), many blue bars have high attributions. The pattern of blue bars does overlap with the pattern of orange bars. It means the ML model based on correct malicious events to make a positive prediction. We believe the ML model relies on correct information for classification and is trustworthy.

\subsection{Adversarial Studies}

Adversarial attacks were done by a simulated ransomware application, the Red team. It can do the core business of ransomware, e.g. rename, delete and encrypt files with several evasive mechanisms. The evasive tricks include changing the file I/O pattern or lower the entropy of encrypt file. I/O pattern changes can be done by actively creating benign events or by slowing down the malicious activities (so  events from other benign processes will go in). Many benign events insertion can change the I/O pattern and dilute the effects of malicious ones. Changing entropy can be done by insert low-entropy data to output file or simply encrypt partial content of the victim files. We found it won't take long for the Red team to find a condition to bypass the detection. To defense the adversarial, the model was probed by the Red team with several different conditions of evasive tricks. All the false negatives were collected for an adversarial re-training. The resiliency of model was improved after several attack-retrain iterations and the Red team can't easily evade the detection. 

\begin{figure}[h]
\centering
\includegraphics[width=.5\textwidth]{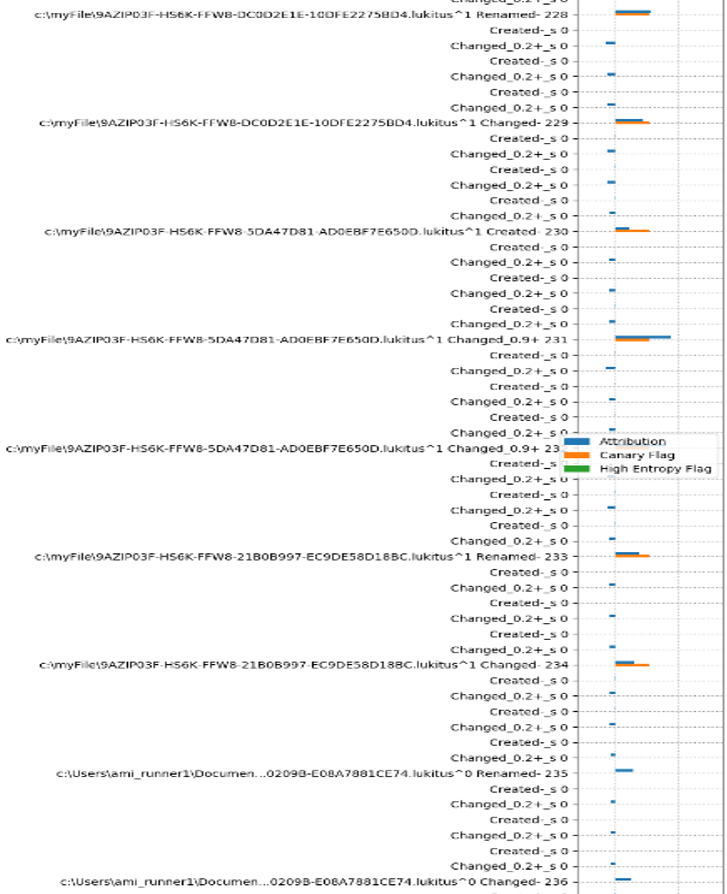}
\caption{\label{fig:ig_advrsrl} The attribution of an adversarial sample log for LSTM model. An adversarial sample was synthesized by inserting six benign I/O events (low attribution events) between malicious events. The orange bar indicates the decoy/malicious events. The attribution amount of decoy event is less than one tenth comparing to the same event from Figure \ref{fig:ig}(b). This sample was predicted as a negative with score 0.01. The original sample without insertion has a positive label with score 0.96} 
\end{figure}

The Integrated Gradient was used to explain how the adversarial samples evade the detection of LSTM model. The synthesized sample, in Figure \ref{fig:ig_advrsrl}, was done by inserting six benign events between malicious events (by method in Figure \ref{fig:data_aug}d). The malicious events indicated by orange bars in Figure \ref{fig:ig_advrsrl} still have positive attributions, but the amount of attribution, the blue bar, is weaken. The total effect of insertion dilutes the malicious events and the prediction was flipped. The dilution effect is not linear. When inserting up to five benign events, the prediction is still positive with a score 0.84. 

The orders of malicious events to LSTM model were also studied. The samples generated by the keyed method, e.g. relocation, reverse and random shuffle (Figure \ref{fig:data_aug}e), were tested . "Relocation" was done by move the second half of log to the front of first half. "Reverse" was reverse the order of all events. All three operations didn't flip the prediction outcome, but they do lower the prediction scores of LSTM from original 0.96 to relocation 0.83, reverse 0.81 and shuffle 0.88. We believe individual events, the order and the pattern of events all contribute to the LSTM decision making.

\section{Discussion}

The bare-metal sandbox system in this paper leverages fast storage interface to promptly refresh the Guest drive back to its original state (within \textasciitilde40 seconds for \textasciitilde30Gb image). The refresh time depends on the I/O speed of storage devise. The system utilizes a common boot loader to programmatically switch booting between Master and Guest OS. Since the Master and Guest OS on different physical storage drives, the Master drive can be protected on the device level by simple mechanisms, e.g. unloading the driver of SSD etc. Without the need to modify storage driver, OS or the VMM for a simulated bare-metal system, the simple implementation is the main advantage of this system. For a large scale malware analysis, the bare-metal machine farm can be a time saving solution. The Control server with programmable power control can detect and force a reboot/power cycle to any non-responsive machines. With small form factor machines, e.g. Intel\textsuperscript{\textregistered} NUCs, a sandbox farm can be built within a limited space.

The dynamic data of crypto-ransomware is difficult to obtain. Around 20\% active samples were identified from \textasciitilde22,000 downloads in our experiment. The paper by Scaife, N \etal \cite{cryptolock} had a similar ratio. However, the crypto-ransomware was reported to be more popular than the locker-ransomware in the review paper by Bander  Al-rimy \etal \cite{Al-rimy2018}. The possible causes of low active ratio could be: 1. the command and control servers (C\&C server) was shutdown or the connection got blocked; 2. the ransomware was designed for specific victims; or 3. the evasive ransomware detect our sandboxes. Based on Figure \ref{fig:starting_time}(b), the slope of green line is still upward near the end of 5 minute region. The ransomware execution time could be extended more to have more active samples. Also even an execution log didn't show any malicious events, we cannot exclude the possibility that our data collection process got killed or blocked before or during ransomware execution. We may investigate further for samples without decoy events or with an uncompleted log if resource available.

The feature selected in ML models are I/O event, path and file entropy. The intuition behind is that file events and encryption are the core malicious activities which would be very unique to crypto-ransomware. We found the ML classifiers can be quickly trained by a SVM or by a simple architecture for LSTM. Categorizing the raw event logs into a series of feature id make the data processing simple and the inference can be done quickly in the detector. Several other ML models such as linear logistic regression or multi-layer perceptron (MLP) etc. were also trained and evaluated. The performances of these models are similar to SVM or LSTM. The path flag added to the feature can help to distinguish the source of I/O event and lower the FPR. However it doubles the feature dimension for LSTM (9 to 18) and more than triple for N-gram linear SVM (90 to 336). To alleviate the concerns of over-fitting, the out-of-sample data from \textasciitilde700 ransomware log were used to evaluate the performance. This out-of-sample dataset was from a newer download. It can be predicted with a good accuracy by a classifier trained by older samples.

In addition to the regular ML pipeline to train and optimize a ML classifier, the block bootstrapping was used to augment the dataset for the needs of the detection. A naive trained model won't perform well in real world especially when the detector need to make decisions based on the partial data extracted from an event stream. The data for a live detector is different from the sample generated directly from the sandbox. Modifying dataset by bootstrapping and oversampling were found to be effective to remove the difference. The classifier trained by augmented dataset was proved to be able to focus on the correct regions of I/O events and doing prediction by limited the amount of data, see Figure \ref{fig:early_wind_test}.

The adversarial research in this paper is done by a grey-box attack. It started from changing the event pattern and the entropy of victim files. The simulated ransomware didn't know the weights or architectures of the model, but do assume the I/O event and entropy of file content were used for featuring. Attacks by changing the I/O event pattern, e.g. slowing down the malicious activities, may have a higher risk for an adversary to be detected. Actively insert benign events to blur malicious ones would be a better option for evasion. 

Figure \ref{fig:ig_advrsrl}, the attribution of a synthesized ransomware log, illustrates the I/O event insertion effects. Inserting low attribution events to the log does blur the attribution of malicious events. Also if an adversary could avoid generating high entropy events, e.g. by a partial encryption etc., the entropy feature  could become futile. Although entropy is an effective trait to detect encryption, it is also an obvious feature for adversary to dodge. To defense adversaries, ML model based on few simple features won't be robust based on our Red team attacking tests. The ideal ML detector should based on multiple featuring methods and inference by an ensemble algorithms. An inference based on collective features increases the depth of defense and could make adversaries more difficult to evade.

Most of existing data augmentation techniques were developed by computer vision community. They were widely used on convolution neural network related applications for image, video and audio datasets. In an image sample, each feature/pixel could be an arbitrary value. But for time series data, such as execution log or API call log, these augmentation techniques did not fit well and may cause the “synthetic gap”. The synthetic gap is the discrepancy between the synthetic and “real” samples. The time series data consist a sequence of observations which may have temporal dependence between them. For example, “write file” event need to go after the “open file” event etc. The generated samples won’t make sense or is “un-real”, if the temporal relationships or important key observation don’t exist. 

The "keyed" method to synthesize new time-series samples was proposed in this paper. It can avoid the “synthetic gap” and the extra filtering. It relies on human to identify a set of “key observations” and preserves the feature constraints before applying data augmentation methods. With the help of domain experts, the heuristics or known golden rules were identified, the “close-to-real” variants can be promptly generated in the data space. Please see Figure \ref{fig:data_aug}.

 \section{Acknowledgements}
 I would like to thank Erdem Aktas, Li Chen and Zheng Zhang for their expert advises and encouragements throughout this project. This project would have been impossible without the pioneer data collection program from Erdem Aktas.

\bibliographystyle{plain}
\bibliography{refs}

\end{document}